\documentclass[runningheads]{llncs}
\usepackage[misc]{ifsym}
\usepackage{cite}
\usepackage{makecell}
\usepackage[shortlabels]{enumitem}
\usepackage{mathtools}
\usepackage{subcaption}
\usepackage{multirow}

\usepackage{cleveref}

\renewcommand{\figurename}{Fig.{}}
\Crefformat{figure}{#2Fig.~#1#3}
\Crefmultiformat{figure}{Figs.~#2#1#3}{ and~#2#1#3}{, #2#1#3}{ and~#2#1#3}
\crefrangeformat{figure}{figs. #3#1#4 to #5#2#6}
\Crefformat{table}{#2Table~#1#3}
\Crefmultiformat{table}{Table~#2#1#3}{ and~#2#1#3}{, #2#1#3}{ and~#2#1#3}
\crefrangeformat{table}{table #3#1#4 to #5#2#6}
\Crefformat{equation}{#2Eq.~(#1#3)}
\crefrangeformat{equation}{eqs.~#3(#1)#4 to #5(#2)#6}
\Crefmultiformat{equation}{Eqs.~#2(#1)#3}{ and~#2(#1)#3}{, #2(#1)#3}{ and~#2(#1)#3}

\Crefformat{algorithm}{#2Algorithm ~#1#3}
\Crefmultiformat{algorithm}{Algorithms ~#2#1#3}{ and~#2#1#3}{, #2#1#3}{ and~#2#1#3}
\crefrangeformat{algorithm}{Algorithms #3(#1)#4 to #5(#2)#6}

\Crefformat{lemma}{#2Lemma ~#1#3}
\Crefmultiformat{lemma}{Lemmas ~#2#1#3}{ and~#2#1#3}{, #2#1#3}{ and~#2#1#3}
\crefrangeformat{lemma}{Lemmas #3#1#4 to #5#2#6}

\Crefformat{section}{#2Section~#1#3}
\Crefmultiformat{section}{Sections~#2#1#3}{ and~#2#1#3}{, #2#1#3}{ and~#2#1#3}
\crefrangeformat{section}{Sections #3#1#4 to #5#2#6}
\Crefformat{subsection}{#2Subsection~#1#3}
\Crefmultiformat{subsection}{Subsections~#2#1#3}{ and~#2#1#3}{, #2#1#3}{ and~#2#1#3}
\crefrangeformat{subsection}{Subsections #3#1#4 to #5#2#6}

\begin{document}
	\title{Double-Wing Mixture of Experts for \\Streaming Recommendations}
	
	\author{Yan Zhao\inst{1 \and 2} \and
		Shoujin Wang\inst{2} \and
		Yan Wang \inst{2} \and Hongwei Liu\inst{1}$^\text{\tiny\Letter}$  \and Weizhe Zhang \inst{1, 3}}
	\authorrunning{Yan Zhao et al.}
	\institute{School of Computer Science and Technology, Harbin Institute of Technology, China \\
		\email{\{yanzhao,liuhw,wzzhang\}@hit.edu.cn}\\
		\and
		Department of Computing, Macquarie University, Australia\\
		\email{\{shoujin.wang,yan.wang\}mq.edu.au}
		\and
	    Cyberspace Security Research Center, Peng Cheng Laboratory, Shenzhen, China\\}
	
	\let\oldmaketitle\maketitle
	\renewcommand{\maketitle}{\oldmaketitle\setcounter{footnote}{0}}
	
	\maketitle
			\vspace{-5mm}
	\begin{abstract}
		Streaming Recommender Systems (SRSs) commonly train recommendation models on newly received data only to address \textit{user preference drift}, i.e., the changing user preferences towards items. However, this practice overlooks the\textit{ long-term user preferences} embedded in historical data. More importantly, the common \textit{heterogeneity} in data stream greatly reduces the accuracy of streaming recommendations. The reason is that different preferences (or characteristics) of different types of users (or items) cannot be well learned by a unified model.
		To address these two issues, we propose a Variational and Reservoir-enhanced Sampling based Double-Wing Mixture of Experts framework, called VRS-DWMoE, to improve the accuracy of streaming recommendations. In VRS-DWMoE, we first devise variational and reservoir-enhanced sampling to wisely complement new data with historical data, and thus address the user preference drift issue while capturing long-term user preferences.
		After that, we propose a Double-Wing Mixture of Experts (DWMoE) model to first effectively learn heterogeneous user preferences and item characteristics, and then make recommendations based on them. 
		Specifically, DWMoE contains two Mixture of Experts	(MoE, an effective ensemble learning model) 
		to learn user preferences and item characteristics, respectively. Moreover, the multiple experts in each MoE learn the preferences (or characteristics) of different types of users (or items) where each expert specializes in one underlying type. Extensive experiments demonstrate that VRS-DWMoE consistently outperforms the state-of-the-art SRSs.
		
	\end{abstract}   
	
		\vspace{-5mm}

	\section{Introduction}
	{\let\thefootnote\relax\footnotetext{This paper was accepted by WISE'2020. The final authenticated publication is available online at https://doi.org/[DOI to be inserted]}}
	\noindent Recommender Systems (RSs) have played an increasingly important role to assist users to make wise decisions. Nowadays, E-commerce platforms generate continuous data stream, e.g., continuous users' purchase records, at an unprecedented speed, which poses new challenges for RSs. Conventional offline RSs train recommendation models with large-volume data periodically~\cite{wwwj_offline1,wwwj_offline2}, and thus cannot process data stream in a real-time manner.
	To this end, \textit{Streaming Recommender Systems} (SRSs) \cite{spmf}, which perform real-time recommendations based on the data stream, have emerged. 
	
	Various SRSs have been proposed through different ways. 
	As an earlier attempt, researchers have constructed SRSs by adapting offline RSs to the streaming setting through training the recommendation models incrementally with new data, e.g., incremental collaborative filtering~\cite{ICF}. Later on, SRSs specifically devised for the streaming scenario have been proposed, e.g., Neural Memory Recommender Networks (NMRN)~\cite{nmrn-gan}.
	
	Despite many SRSs have been proposed, the following two challenges still need to be well addressed to improve the accuracy of streaming recommendations:
	\textbf{CH1:} how to address \textit{user preference drift}, i.e., the user preferences towards items changing over time~\cite{wwwj_user_preference_drift}, while capturing \textit{long-term user preferences}, and \textbf{CH2:} how to handle the \textit{heterogeneity} of users and items, i.e., different types of users (or items) have different preferences (or characteristics).
	To address \textbf{CH1}, Wang et al.~\cite{nmrn-gan} and Wang et al.~\cite{spmf} have proposed a neural memory network based approach, i.e., NMRN, and a reservoir-based approach, i.e., Stream-centered Probabilistic Matrix Factorization (SPMF), respectively. However, NMRN has difficulties in capturing long-term user preferences as the preferences stored in memory might be overwritten frequently over the continuous data stream, while SPMF has limited capability to address user preference drift as it could not sufficiently learn from new data to capture the changing user preferences. Different from the first challenge, \textbf{CH2} has not been discussed in the literature of SRSs. Existing SRSs commonly utilize a unified model to learn user preferences and item characteristics for all users and items~\cite{ssrm,eals}. However, they cannot well deal with the intrinsic difference between user preferences and item characteristics. Moreover, the preferences (or characteristics) of different types of users (or items) cannot be well learned by a unified model, either.
	
	\noindent\textbf{Our Approach and Contributions}. To address the above two challenges, we propose a novel Variational and Reservoir-enhanced Sampling based Double-Wing Mixture of Experts framework, called VRS-DWMoE, to improve the accuracy of streaming recommendations. Specifically, VRS-DWMoE contains two key components: 1) Variational and Reservoir-enhanced Sampling (VRS), which wisely samples historical data containing long-term user perferences from the reservoir, i.e., a set of representative historical data, with an adjustable sampling size to complement new data, and 2) Double-Wing Mixture of Experts (DWMoE), which first learns heterogeneous user preferences and item characteristics with the training data prepared by VRS, and then utilizes the learned preferences and characteristics to make recommendations. Specifically, DWMoE contains two  elaborately devised Mixture of Experts (MoEs) to learn user preferences and item characteristics, respectively. Note that MoE is an effective ensemble learning model which wisely fuses the outputs of multiple experts, i.e., atomic models specializing in different types of input, for better learning performance~\cite{moe_survey}. Moreover, the multiple experts in each of the aforementioned two MoEs learn the preferences (or characteristics) of different types of users (or items) where each expert specializes in one underlying type.
	
	The characteristics and contributions of our work are summarized as follows:
	\begin{itemize}
		\item In this paper, we propose a novel VRS-DWMoE framework, which consists of Variational and Reservoir-enhanced Sampling (VRS) and Double-Wing Mixture of Experts (DWMoE), for accurate streaming recommendations.
		\item To address \textbf{CH1}, we propose VRS to wisely complement new data with historical data while guaranteeing the proportion of new data. In this way, VRS not only captures long-term user preferences from the sampled historical data, but also effectively addresses user preference drift by highlighting the importance of new data. 
		\item To address \textbf{CH2}, we propose DWMoE to first effectively  learn heterogeneous user preferences and item characteristics, and then make recommendations with learned preferences and characteristics. Specifically, DWMoE not only learns user preferences and item characteristics with two elaborately devised MoEs, respectively, to deal with their intrinsic difference, but also allows each expert to specialize in one underlying type of users (or items) to more effectively learn the heterogeneous user preferences (or item characteristics). 
	\end{itemize}
	\vspace{-3mm}
	\section{Related Work}
	\vspace{-2mm}
	In this section, we first review streaming recommender systems, and then introduce mixture of experts, based on which we propose VRS-DWMoE.
	
	\vspace{-3mm}
	\subsection{Streaming Recommender Systems}
	
	The early SRSs enhance offline RSs with elaborately devised online update mechanisms for streaming recommendations. For example, Papagelis et al.~\cite{ICF} adapt user-based collaborative filtering to the streaming setting by incrementally updating the user-to-user similarities.
	Later on, several approaches have been proposed to adapt the matrix factorization to the streaming setting, including incremental stochastic gradient descent~\cite{ISGD}, randomized block coordinate descent~\cite{rcd}, and fast alternating least square~\cite{eals}. Moreover, the optimization methods for offline matrix factorization have also been applied to the streaming setting, including pair-wise personalized ranking~\cite{rmfk} and Bayesian inference~\cite{vbmf}. 
	
	Recently, SRSs focusing on the challenges in streaming recommendations have emerged. To address user preference drift while capturing long-term user preferences, the neural memory network based approach, i.e., NMRN~\cite{nmrn-gan}, and the reservoir-based approach, i.e., SPMF~\cite{spmf}, have been proposed.  Specifically, NMRN tries to capture user preferences by neural memory network. However, it has limited capability to capture long-term user preferences, as the long-term preferences stored in the memory might be overwritten frequently over the continuous data stream. In addition, SPMF maintains a reservoir and trains the recommendation model with both sampled historical data and sampled new data. However, SPMF has difficulties in addressing user preference drift, as it equally treats the historical data and new data when conducting the sampling process and thus overlooks the importance of new data. 
	In addition, to avoid the limitations of a single model, OCFIF~\cite{ocfif}  ensembles multiple recommendation models to conduct streaming recommendations. However, it does not fully exploit the potential of these recommendation models, as it trains the recommendation models independently and selects only one model for recommendations. 
	
	\noindent \textbf{Summary.} More work is needed to address user preference drift while capturing long-term user preferences.  In addition, studies on addressing the issue of user or item heterogeneity in streaming recommendations have not been reported.
	
	\subsection{Mixture of Experts}
	Mixture of Experts (MoE)~\cite{moe_survey} is an effective ensemble learning model which wisely fuses the results of multiple experts to achieve better learning performance. Specifically, MoE contains 1) multiple experts where each expert is an atomic model specializing in learning from a particular type of input data, 2) a gating network to calculate the gating weights, i.e., the expertise of each expert regarding the input, and 3) a fusion module to fuse the outputs of the experts with the gating weights. Its effectiveness has been verified in various areas~\cite{moe_nlp}, including offline recommendations~\cite{moe_recommendation_multitask}.
	However, these existing MoE based RSs commonly employ MoE to perform multi-task learning~\cite{moe_recommendation_multitask} or simply combine multiple independent RSs~\cite{moe_recommendation_combine}, rather than effectively learning the heterogeneous user preferences and  item characteristics from the interactions. In addition, although MoE has achieved good performance regarding streaming data in multiple areas~\cite{moe_streaming_robotic_system},
	its effectiveness in streaming recommendations has not been explored.
	
	\noindent \textbf{Summary.} The potential of MoE has not been explored by existing SRSs. Although MoE based offline RSs have been proposed, they cannot well learn the heterogeneous user preferences and item characteristics. Moreover, conventional MoE based approaches only contain a single MoE, and thus can not be employed directly to well address the aforementioned challenge, i.e., CH2.

	\vspace{-1.5mm}
	\section{VRS-DWMoE Framework}
	\vspace{-1.2mm}
	
	We first formulate our research problem. After that, we propose Variational and Reservoir-enhanced Sampling based Double-Wing Mixture of Experts (VRS-DWMoE), and then introduce its two key components, i.e., VRS and DWMoE.

	\vspace{-1.3mm}
	\subsection{Problem Statement}
	
	In this section, we formulate the research problem of streaming recommendations with implicit interactions, e.g., the users' purchase records of items. Given the user set $\mathbf{U}$ and item set $\mathbf{V}$, we use $y_{u,v}$ to denote an interaction between user $u \in \mathbf{U}$ and item $v \in \mathbf{V}$. Then, the list of currently received interactions is denoted by $\mathbf{Y} = \{ y^1_{u^1,v^1}, y^2_{u^2,v^2}, \dots, y^k_{u^k,v^k},$ $\dots \}$. Note that the interactions in $\mathbf{Y}$ are sorted based on their receiving time, e.g., $y^k_{u^k,v^k}$ indicates the $k^\text{th}$ received interaction. In addition, as in the real-world data stream, the adjacent interactions, e.g., $y^k_{u^k,v^k}$ and $y^{k+1}_{u^{k+1},v^{k+1}}$, may involve different users, i.e., user $u^k$ is possibly different from user $u^{k+1}$. With the above information, given the target user $u'$ and target item $v'$, the task of SRSs can be formulated as $\hat{y}_{u',v'} = P(y_{u',v'}|\mathbf{Y})$, i.e., predicting the probability of an interaction between the target user and target item conditional on the currently received interactions.
	\vspace{-2.2mm}
	%		\vspace{-mm}
	\subsection{Our Proposed VRS-DWMoE Framework}
	\begin{figure}[t]
		
		\centering
		\includegraphics[width= 0.9\textwidth, keepaspectratio]{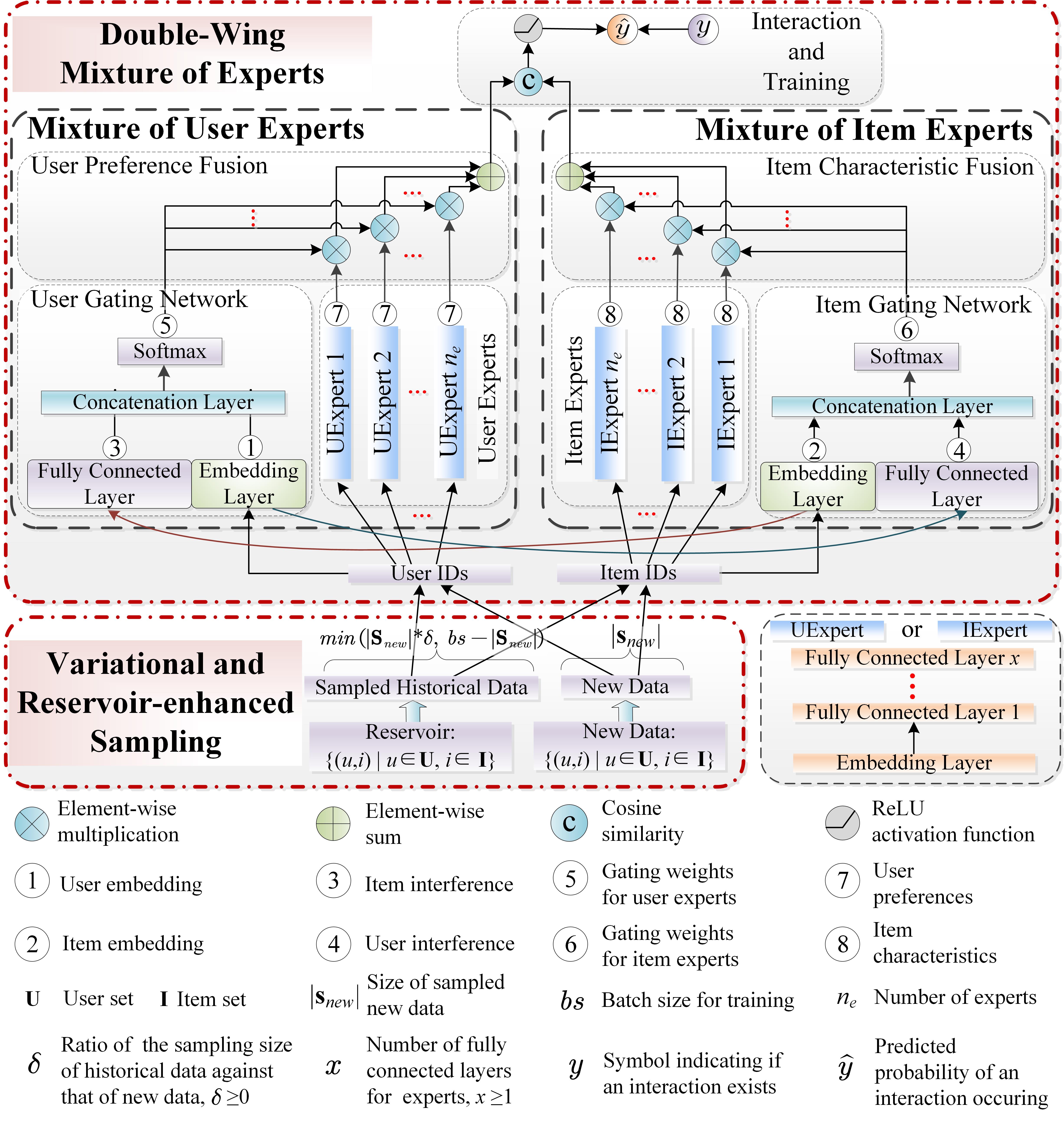}
		\vspace{-3mm}	
		\caption{Workflow of our proposed VRS-DWMoE.}
		\vspace{-3mm}
		\label{figure:architecture} 
	\end{figure}
	To simultaneously address user preference drift while capturing long-term user preferences and handle the heterogeneity of users and items, we propose VRS-DWMoE, which contains two key components: 1) Variational and Reservoir-enhanced Sampling (VRS), and 2) Double-Wing Mixture of Experts (DWMoE). 
	Specifically, as shown in~\Cref{figure:architecture}, VRS first complements the new data with sampled historical data while guaranteeing the proportion of new data in preparation for training. After that, with the training data prepared by VRS, DWMoE better learns heterogeneous user preferences and item characteristics with two elaborately devised Mixture of Experts (MoEs), i.e., Mixture of User Experts (MoUE) and Mixture of Item Experts (MoIE), respectively, and then makes recommendations based on the learned user preferences and item characteristics.  
	\vspace{-4mm}
	\subsection{Variational and Reservoir-enhanced Sampling}
	\label{ssec:vrs}
	The continuous and infinite data stream makes it impractical for SRSs to train recommendation models with all the data. To this end, we propose VRS to prepare the data for training by wisely complementing new data with sampled historical data while guaranteeing the proportion of new data.
	
	Specifically, following~\cite{reservoir-boosting,spmf}, we first maintain a reservoir to store a set of representative historical data. As newer data commonly reflect more recent user preferences, we put new data into the reservoir and discard the oldest data when the reservoir runs out of space. With this reservoir and new data, VRS generates the training data with two different strategies in two typical scenarios for streaming recommendations, i.e., the \textit{underload} scenario, where the data receiving speed is lower than the data processing speed, and the \textit{overload} scenario, where the data receiving speed is higher than the data processing speed, respectively. Note that the contribution of VRS mainly lies in the underload scenario, which is more common in the real world~\cite{underload_forbes}. More details are presented below.
	
	\noindent\textbf{Underload Scenario}. In the underload scenario, VRS generates training data by first sampling representative historical data $\textbf{S}_{his}$ from the reservoir with a variational sampling size, and then merging $\textbf{S}_{his}$ and all the new data \textbf{N} (i.e., all the new data are sampled by VRS in the underload scenario). Specifically, with the training batch size $bs$ and a predetermined parameter $\delta$ ($\delta \geq 0$) measuring the ratio of the sampling size $|\textbf{S}_{his}|$ of reservoir against the size $s_{new}$ of new data, the sampling size $|\textbf{S}_{his}|$of reservoir can be calculated as below,
	\begin{equation}
	\label{eq:samping_size}
	|\textbf{S}_{his}| = min(s_{new}*\delta, bs-s_{new}).
	\end{equation}
	In this way, the sampling size of reservoir can be adjusted by $\delta$ based on the characteristics of data stream. For example, $\delta$ should be set to a small value, e.g., 0.1, for data stream where users' preferences change frequently to focus more on new data. Then, $\textbf{S}_{his}$ is sampled from the reservoir based on their receiving time, i.e., more recent received interactions are assigned higher sampling probabilities. Specifically, to reflect the importance of newer data, we employ a decay ratio $\lambda_{res}$ $(\lambda_{res} > 1)$ to assign higher sampling probabilities to newer data in the reservoir,
	\begin{equation}
	\label{eq:recursion}
	p_k = p_{k-1} * \lambda_{res},
	\end{equation}
	where $p_i$ denotes the sampling probability of the $i^{th}$ received interaction. Assuming that the sampling probability of the earliest received interaction in the reservoir is $p_1$, we can get $p_k$ by iteratively performing \Cref{eq:recursion}, i.e.,
	\begin{equation}
	\label{eq:pk}
	p_k = p_{1} * (\lambda_{res})^{k-1}.
	\end{equation}
	Then, with~\Cref{eq:pk}, we can obtain the normalized sampling probabilities, taking the probability of the $k^{th}$ received interaction as an example,
	
	\begin{equation}
	\label{eq:probability}
	P(k | \lambda_{res}, s_{res}) = \frac{p_k}{\sum_{i=1}^{s_{res}} p_i}=\frac{(\lambda_{res})^{k-1} * (1-\lambda_{res})}{1-(\lambda_{res})^{s_{res}}},
	\end{equation}
	where $s_{res}$ denotes the size of the reservoir.
	
	With the normalized sampling probabilities and the sampling size calculated in~\Cref{eq:samping_size}, VRS samples representative historical data $\textbf{S}_{his}$ from the reservoir. Finally, the training data \textbf{T} is obtained by merging  $\textbf{S}_{his}$ and the new data $\textbf{N}$.
	
	\noindent\textbf{Overload Scenario}. In the overload scenario, VRS only samples $\textbf{S}_{new}$  from new data to form the training data, i.e., $|\textbf{S}_{his}|$ is set to 0, for effectively capturing the latest user preferences. The sampling probability of the new data can be calculated in a similar way as described by~\Cref{eq:recursion,eq:pk,eq:probability}, i.e.,
	\begin{equation}
	P(k |\lambda_{new}, s_{new}) = \frac{p_k}{\sum_{i=1}^{s_{new}} p_i}=\frac{(\lambda_{new})^{k-1} * (1-\lambda_{new})}{1-(\lambda_{new})^{s_{new}}},
	\end{equation}
	where $\lambda_{new}$ and $s_{new}$ denote the decay ratio and size, respectively, of new data. With this sampling probability and the sampling size set to be the batch size $bs$, VRS samples $\textbf{S}_{new}$ from the new data as the training data \textbf{T}. Note that, in the case where the data receiving speed exactly equals to the data processing speed, VRS utilizes the entire new data to form the training data \textbf{T}.
	
	\subsection{Double-Wing Mixture of Experts}
	\label{ssec:ael}
	With the training data \textbf{T} prepared by VRS, 
	DWMoE first utilizes two MoEs, i.e., MoUE and MoIE, to learn user preferences and item characteristics, respectively, to deal with their intrinsic difference~\cite{wang2020transaction}. Moreover, each expert in MoUE (or MoIE) specializes in one underlying type of users (or items) for more effectively learning heterogeneous user preferences (or item characteristics). Then, DWMoE makes recommendations with the learned user preferences and item characteristics.
	
	Specifically, MoUE and MoIE share the same structure with different parameters. This structure has three key parts: 1) multiple experts, 2) a gating network, and 3) a fusion module. Taking MoUE as an example, multiple experts first learn the user preferences in parallel. Then, the gating weights, which measure the expertise of each expert regarding each input user, are calculated by the gating network. After that, the fusion module calculates the unified preferences for each user by fusing the preferences learned by all experts with the gating weights. Note that we set the numbers ($n_e$) of experts in MoUE and MoIE the same in this paper, and will study the effect of different numbers of experts in MoUE and MoIE in the future work.
	More details are presented below.

	\noindent\textbf{Expert}. The experts in MoUE and MoIE learn user preferences and item characteristics, respectively. Taking MoUE as an example, 
	each expert first utilizes an embedding layer 
	to learn the user embedding $\textbf{p}_u^i$, where $i$ denotes the index of the expert. Then, the user preferences are learned by the experts with $x$ ($x \geq 1$) fully connected layers, taking the user preferences $\textbf{P}^{i}_u$ as an example,
	\begin{equation}
	\small
	\label{eq:MoUE_expert}
	\textbf{P}^{i}_u = a^{MoUE}_{i,x}(\cdots a^{MoUE}_{i,2}(\textbf{W}^{MoUE}_{i,2}a^{MoUE}_{i,1}(\textbf{W}^{MoUE}_{i,1}\textbf{p}_u^{i} + \textbf{b}^{MoUE}_{i,1})+\textbf{b}^{MoUE}_{i,2})\cdots),
	\end{equation}
	where $a^*_*$, $\textbf{W}^*_*$, and $\textbf{b}^*_*$ denote the activation function, weight matrix, and bias vector, respectively. For example, $a^{MoUE}_{i,x}$, $\textbf{W}^{MoUE}_{i,x}$, and $\textbf{b}^{MoUE}_{i,x}$ denote the activation function, weight matrix, and bias vector, respectively, in the $x^{th}$ layer for the $i^{th}$ expert in MoUE. Note that the symbols $a^*_*$, $\textbf{W}^*_*$, and $\textbf{b}^*_*$ are used in the rest of this paper with different superscripts and subscripts to introduce the fully connected layers. In a similar way as described by \Cref{eq:MoUE_expert}, item characteristics $\textbf{Q}^{j}_v$ can be learned by the $j^{th}$ expert in MoIE with the item embedding $\textbf{q}_v^{j}$,
	\begin{equation}
	\textbf{Q}^{j}_v = a^{MoIE}_{j,x}(\cdots a^{MoIE}_{j,2}(\textbf{W}^{MoIE}_{j,2}a^{MoIE}_{j,1}(\textbf{W}^{MoIE}_{j,1}\textbf{q}_v^{j}+\textbf{b}^{MoIE}_{j,1}) + \textbf{b}^{MoIE}_{j,2})\cdots).
	\end{equation}
	
	\noindent\textbf{Gating Network}. The gating networks in MoUE and MoIE calculate the gating weights measuring the expertise scales of each expert in learning the preferences of input users and characteristics of input items,
	respectively. To achieve this goal, taking the gating network in MoUE as an example, we first employ an embedding layer and a fully connected layer to calculate user embedding $\textbf{p}_u^{MoUE}$ and item interference $\textbf{I}^{MoUE}$, respectively. Note that the item interference is employed to more accurately calculate the gating weights in MoUE by taking the items interacted with the corresponding users into consideration. Specifically, the item interference can be calculated with the item embedding $\textbf{q}_v^{MoIE}$ in MoIE,
	\begin{equation}
	\label{eq:interference}
	\textbf{I}^{MoUE} = a_{gate}^{MoUE}({\textbf{W}}^{MoUE}_{inter}\textbf{q}_v^{MoIE} + {\textbf{b}}^{MoUE}_{inter}).
	\end{equation}
	Then, the user embedding and item interference are concatenated, 
	\begin{equation}
	\label{eq:concatenation}
	\textbf{c}^{MoUE} = \begin{bmatrix} 
	\textbf{p}_u^{MoUE}; 
	\textbf{I}^{MoUE}    
	\end{bmatrix}.
	\end{equation}
	After that, $\textbf{c}^{MoUE}$ is fed into a softmax layer to get the gating weights $\textbf{g}^{MoUE}$,
	\begin{equation}
	\label{eq:softmax}
	\textbf{g}^{MoUE} =softmax(\textbf{W}^{MoUE}_{soft}\textbf{c}^{MoUE} + \textbf{b}^{MoUE}_{soft}).
	\end{equation}
	Likewise, the gating weights $\textbf{g}^{MoIE}$ for the experts in MoIE can be calculated in a similar way as described by \Cref{eq:interference,eq:concatenation,eq:softmax}.
	
	\noindent\textbf{Fusion Module}. The user preferences (or item characteristics) learned by multiple experts in MoUE (or MoIE) are fused to the unified ones to fully utilize the expertise of all the experts. Specifically, with the above calculated user preferences, item characteristics, and their corresponding gating weights, the unified user preferences $\textbf{P}_u^{uni}$ and unified item characteristics $\textbf{Q}_v^{uni}$ can be calculated with the dot production, respectively,
	\begin{equation}
	\textbf{P}_u^{uni} = [\textbf{P}^{1}_u;\cdots;\textbf{P}^{n_e}_u]^T\textbf{g}^{MoUE},
	\end{equation}
	\begin{equation}
	\textbf{Q}_v^{uni} = [\textbf{Q}^{1}_v;\cdots;\textbf{Q}^{n_e}_v]^T\textbf{g}^{MoIE},
	\end{equation}
	where $n_e$ denotes the number of experts.
	
	\noindent\textbf{Interaction Module}. To make recommendations based on unified user preferences and unified item characteristics, we first utilize cosine similarity to measure how the unified preferences match the corresponding unified characteristics, %more space, unimportant reference
	\begin{equation}
	cos\_sim_{<\textbf{P}_u^{uni},\textbf{Q}_v^{uni}>} = cosine(\textbf{P}_u^{uni},\textbf{Q}_v^{uni}) = \frac{(\textbf{P}_u^{uni})^T  \textbf{Q}_v^{uni}}{\|\textbf{P}_u^{uni}\|_2 \|\textbf{Q}_v^{uni}\|_2}, \label{eq:sim} \\
	\end{equation}
	and then we obtain the predicted probability $\hat{y}_{u,v}$ of the interaction between user $u$ and item $v$ by performing a nonlinear transformation of this cosine similarity,
	\begin{equation}
	\hat{y}_{u,v} = a_{out}^{predict}({\textbf{W}}_{out}^{predict}	cos\_sim_{<\mathbf{P}_u^{uni},\mathbf{Q}_v^{uni}>} + \textbf{b}_{out}^{predict}).
	\end{equation}
	%	where $a_{out}^{predict}$, $\textbf{W}^{predict}$ $\textbf{b}^{predict}$ denote the activation function, weight matrix, and bias vector, respectively.

	\noindent\textbf{Optimization}. To learn the parameters of our proposed DWMoE, we train the model by minimizing the following loss with stochastic gradient descent,
	\begin{equation}
	\ell oss^{total} = \ell oss^{acc} + \gamma(\ell oss^{gate}),
	\end{equation}
	where $\ell oss^{acc}$ is the loss for the recommendation accuracy, $\ell oss^{gate}$ is used as the regularization term for gating weights to avoid local optimizations, and $\gamma$ is the coefficient to adjust the importance of $\ell oss^{gate}$.
	
	Specifically, to measure the difference between the ground truth and the prediction, we employ the cross-entropy loss as below,
	\begin{equation}
	\ell oss^{acc}(y_{u,v},\hat{y}_{u,v}) = -(y_{u,v} log (\hat{y}_{u,v}) + (1-y_{u,v})log (1-\hat{y}_{u,v})),
	\end{equation}
	where $y_{u,v}$ is the label of the interaction between user $u$ and item $v$, i.e., it is 1 if this interaction exists and 0 otherwise, and $\hat{y}_{u,v}$ denotes the predicted probability for this interaction. Moreover, we introduce $\ell oss^{gate}$ to avoid the local optimization caused by the imbalanced utilization of experts, i.e., some experts receive large gating weights for most interactions while others always receive small gating weights. Specifically, we employ the standard derivations of gating weights in both MoUE and MoIE to form  $\ell oss^{gate}$,
	\begin{equation}
	%	\small
	\begin{split}
	&\ell oss^{gate} = \ell oss^{gate}_{MoUE} + \ell oss^{gate}_{MoIE} \\
	& =  \sqrt{\frac{1}{n_e}\sum\nolimits_{i=1}^ {n_e}(\textbf{g}^{MoUE}_i - \bar{\textbf{g}}^{MoUE})^2} +  \sqrt{\frac{ 1}{n_e}\sum\nolimits_{j=1}^{n_e}(\textbf{g}^{MoIE}_j - \bar{\textbf{g}}^{MoIE})^2},
	\end{split}
	\end{equation}
	where $\textbf{g}^{MoUE}_i$ and $\textbf{g}^{MoIE}_j$ denote the gating weight for the $i^{th}$ expert in MoUE and the gating weight for the $j^{th}$ expert in MoIE, respectively, and $\bar{\textbf{g}}^{MoUE}$ and $\bar{\textbf{g}}^{MoIE}$ denote the average of gating weights for MoUE and MoIE, respectively. Through minimizing $\ell oss^{gate}$, DWMoE encourages MoUE and MoIE to more effectively utilize all their experts to learn the heterogeneous user preferences and item characteristics, respectively, and thus to increase the accuracy of streaming recommendations .
	
	\section{Experiments}
	In this section, we present the results of the  extensive experiments we conducted which aim to answer the following four research questions:
	\begin{enumerate}[start=1,label={\upshape\bfseries RQ\arabic*.},wide = 0pt, leftmargin = 3.2em]
		
		\item How does our proposed VRS-DWMoE perform when compared with the state-of-the-art approaches? 
		\item How does our proposed DWMoE perform when compared with the existing recommendation models?
		\item How does our proposed VRS perform when compared with the existing sampling methods?
		\item How does the number of experts in VRS-DWMoE affect the recommendation accuracy?

	\end{enumerate}

	\subsection{Experimental Settings}
	\begin{table}[t]
		\centering
		\caption[...]{Statistics of three datasets used in our experiments\footnotemark.}
		\setlength{\tabcolsep}{8pt}
		\begin{tabular}{|c|c|c|c|c|}
			\hline
			\textbf{Datasets} & \textbf{\#Users} & \textbf{\#Items} & \textbf{\#Interactions} & \textbf{Sparsity} \\ \hline
			MovieLens         & 6400             & 3703             & 994169                  & 95.81\%           \\ \hline
			Netflix           & 5000             & 16073            & 1010588                 & 98.74\%           \\ \hline
			Yelp              & 25677            & 25815            & 731671                  & 99.89\%           \\ \hline
		\end{tabular}
		
		\label{table:datasets}
	\end{table}
	\footnotetext{The symbol \# indicates the number, e.g., \#Users indicates the number of users.}
	\textbf{Datasets}. In the experiments, we employ three widely-used real-world datasets~\cite{spmf,eals}, i.e., MovieLens (1M)\footnote{https://grouplens.org/datasets/movielens/1m}, Netflix\footnote{https://www.kaggle.com/netflix-inc/netflix-prize-data}, and Yelp\footnote{https://www.yelp.com/dataset/challenge}, to verify the effectiveness of our proposed VRS-DWMoE. Note that we extract the interactions of randomly selected 5000 users from the Netflix dataset for the experiments, as processing the original Netflix dataset, which contains more than 100 million interactions, is beyond our computational capacity. In addition, following the common practice~\cite{eals,bpr}, for each dataset, we retain the interactions from users who have more than 10 interactions to reduce data sparsity. 
	The statistics of the tuned datasets are summarized in~\Cref{table:datasets}. 
	Moreover, following~\cite{ocfif,nmrn-gan}, we transform the explicit ratings in all three datasets into the implicit ones, where it is 1 if an explicit rating exists and 0 otherwise, as this work focuses on the recommendations with implicit interactions.
	
	\noindent \textbf{Evaluation Policy}. Following~\cite{eals}, we first sort the data in each of the three datasets by their receiving time, and then divide them into two parts, i.e., 1) the training set to simulate the historical data, and 2) the test set to simulate the upcoming data in the streaming scenario. Specifically, the data in the training set are used for incremental training while the data in the test set are first used for the test and then used for incremental training. We have set the proportion of training set to 85\%, 90\% and 95\%, respectively, for evaluating the performance of our proposed VRS-DWMoE. Due to the space limit, we report the results in the case where the proportion of the training set is 90\% only, as the work in~\cite{eals} does, while the results in the other two cases are similar to the reported ones. 
	Moreover, to verify the effectiveness of our proposed VRS-DWMoE in the underload scenario and overload scenario, we train the recommendation model with a fixed number $n_p$ ($n_p = 256$ in this paper) of interactions each time and adjust the number $n_r$ of interactions received in this training period to indicate different workload intensities. For the sake of simplicity, we use $n_p$ and $n_r$ to simulate the data processing speed $s_p$ and data receiving speed $s_r$, respectively. In this way, the underload scenario and the overload scenario can be simulated by the cases where  $s_p > s_r$ and $s_p < s_r$, respectively. \\
	\textbf{Evaluation Metrics}. Following the common practice~\cite{spmf,eals}, we adopt the ranking-based evaluation strategy. Specifically, for each interaction between a target user and a target item, we first randomly sample 99 items which are not interacted with this user as negative items, and then rank the target item among these 100 items, i.e., the target one plus the 99 sampled ones. Finally, the recommendation accuracy is measured by two widely used metrics, i.e., Hit Ratio (\textbf{HR}) and Normalized Discounted Cumulative Gain (\textbf{NDCG})~\cite{eals,spmf}.\\
	\noindent\textbf{Baselines}. The following eight baselines are used for comparisons, including iBPR, iGMF, iMLP, iNeuMF,  RCD, eAls, SPMF, and OCFIF.  
	\begin{itemize}
		\item \textit{Bayesian Personalized Ranking} (BPR)~\cite{bpr} is a representative personalized ranking method to optimize the matrix factorization. We adapt BPR to the streaming setting, named as \textbf{iBPR}, by training it with new data continuously via stochastic gradient descent.
		\item \textit{Neural Matrix Factorization} (NeuMF)~\cite{ncf} is an advanced matrix factorization model, which combines two other recommendation models, i.e., \textit{Generalized Matrix Factorization} (GMF) and \textit{Multi-Layer Perceptron} (MLP), to achieve higher recommendation accuracy.  We adapt these three recommendation models, i.e., NeuMF, GMF, and MLP, to the streaming setting, named as \textbf{iNeuMF}, \textbf{iGMF}, and \textbf{iMLP}, respectively, by training the recommendation models with new data continuously via stochastic gradient descent.
		\item \textit{Randomized block Coordinate Descent} (\textbf{RCD})~\cite{rcd} and \textit{Element-wise Alternating Least Squares} (\textbf{eAls})~\cite{eals} are two representative approaches for optimizing the matrix factorization models in the streaming setting. We enhance RCD and eAls with abilities of batch processing to increase their throughput for fair comparisons.
		\item \textit{Stream-centered Probabilistic Matrix Factorization} (\textbf{SPMF})~\cite{spmf} is a state-of-the-art SRS. SPMF is originally performed along with a time-consuming sampling method and does not perform well with our evaluation policy where sampling needs to be frequently performed. For a fair comparison, we employ our proposed VRS to prepare training data for SPMF.
		
		\item \textit{Online Collaborative Filtering with Implicit Feedback} (\textbf{OCFIF})~\cite{ocfif} is the only reported SRS employing multiple models (i.e., matrix factorization) to avoid the limitations of a single model for higher recommendation accuracy. 
	\end{itemize}
	In addition, we equip our proposed \textbf{VRS-DWMoE} with different numbers of experts  (i.e., 2, 4, 6, and 8) for comparisons. For example, VRS-DWMoE\_8 indicates VRS-DWMoE equipped with 8 experts for both MoUE and MoIE.

	\noindent\textbf{Parameter Setting}. For a fair comparison, we initialize the baselines with parameters reported in their papers and optimize them for our settings. 
	For our VRS-DWMoE, 
	we empirically set the learning rate to 0.001, the batch size $bs$ to 256, the loss coefficient $\gamma$ to 0.01, and the volume of reservoir to 10000 interactions. Besides, we employ the widely used negative sampling technique~\cite{ocfif,wang2019modeling,wwwj_negative_sampling}, where the reservoir is used to check if an interaction exists and the negative sampling size is set to four, to improve the learning performance. We also adopt L2 regularization and Adam optimizer to avoid overfitting and for the optimization purpose, respectively. Other parameters including $\delta$,  $\lambda_{res}$, and $\lambda_{new}$ are adjusted via cross validation to achieve the best performance in different cases.
	\subsection{Performance Comparison and Analysis}
	\begin{table}[t]
		\caption{Performance comparison with baselines.}
		\centering
		\fontsize{8.8pt}{9.8pt}\selectfont{
			\begin{tabular}{c|c|c|ccc|ccc} 
				\hline
				\multirow{2}{*}{Datasets}
				&\multicolumn{2}{c|}{Metrics}                                                                                      & \multicolumn{3}{c|}{HR@10}                                                                                   & \multicolumn{3}{c}{NDCG@10}                                                          \\ \cline{2-9}
				&\multicolumn{2}{c|}{Data Receiving Speeds ($s_r$)}                                                                        & 128                                & 256                                & 512                                & 128                                & 256                                & 512                                \\ \hline
				\multirow{10}{*}{MovieLens}
				&\multirow{9}{*}{Baselines}                                               &  eAls    & 0.231                              & 0.231                              & 0.234                              & 0.106                              & 0.106                              & 0.108                              \\
				&& RCD    
				& 0.287                              & 0.297                              & 0.278                              & 0.139                              & 0.145                              & 0.138                                 \\
				&& iBPR   
				& 0.303                              & 0.303                              & 0.279                              & 0.147                              & 0.147                              & 0.134        \\
				%& iTPMF-CF  	
				%& 0.408                              & 0.401                              & 0.426                              & 0.222                              & 0.216                              & 0.229        \\
				&& SPMF   
				& 0.460                              & 0.453                              & 0.440                              & 0.251                              & 0.246                              & 0.238                                 \\
				&& iGMF    
				& 0.525                			 & 0.529                              & 0.477                              & 0.295                              & 0.297                              & 0.265                                 \\
				&& iMLP    
				& 0.538                              & 0.539                              & 0.488                              & 0.304                              & 0.303                              & 0.272                                          \\
				&& iNeuMF  
				& \underline{0.551} & \underline{0.546} & \underline{0.496} & \underline{0.311} & \underline{0.307} & \underline{0.275}  \\ 
				&& OCFIF   
				& 0.532                              & 0.508                              & 0.467                              & 0.291                              & 0.279                              & 0.256                                  \\ \cline{2-9}
				&Ours 
				&VRS-DWMoE\_8 
				& \textbf{0.563}    & \textbf{0.558}    & \textbf{0.535}    & \textbf{0.317}    & \textbf{0.313}    & \textbf{0.299}     \\ \cline{2-9}
				\rule{0pt}{8pt} &\multicolumn{2}{c|}{Improvement percentages\footnotemark }                                                              
				& 2.20\%                             & 2.20\%                             & 7.90\%                             & 1.90\%                             & 2.00\%                             & 8.70\%                                \\ \hline \hline
				\multirow{10}{*}{Netflix}
				&\multirow{9}{*}{Baselines}                                               
				& eAls    &  0.395                              & 0.389                              & 0.362                              & 0.211                              & 0.207                              & 0.192                               \\
				&& RCD     & 
				0.447                              & 0.436                              & 0.435                              & 0.226                              & 0.226                              & 0.219                              \\
				&& iBPR    & 
				0.685                              & 0.686                              & 0.627                              & 0.396                              & 0.395                              & 0.360     \\
				&& SPMF    & 
				0.701                              & 0.669                              & 0.640                              & 0.425                              & 0.397                              & 0.378                                    \\
				&& iGMF    & 
				0.747                              & 0.748                              & 0.577                              & 0.482                              & 0.482                              & 0.352                                 \\
				&& iMLP    & 
				0.787                              & 0.782                              & 0.624                              & 0.519                              & 0.510                              & 0.369                                  \\
				&& iNeuMF  & 
				\underline{0.801} & \underline{0.798} & \underline{0.711} & \underline{0.531} & \underline{0.529} & \underline{0.430} \\ 
				&& OCFIF   & 
				0.745                              & 0.734                              & 0.606                              & 0.457                              & 0.453                              & 0.357                                  \\ \cline{2-9}
				&Ours 
				&VRS-DWMoE\_8 
				& \textbf{0.821}    & \textbf{0.814}    & \textbf{0.790}    & \textbf{0.553}    & \textbf{0.548}    & \textbf{0.515}     \\ \cline{2-9}
				\rule{0pt}{8pt}&\multicolumn{2}{c|}{\makecell{Improvement percentages\textsuperscript{5}}} 
				& 2.50\%                             & 2.00\%                             & 11.1\%                             & 4.10\%                             & 3.60\%                             & 19.8\%                                \\ \hline \hline
				\multirow{10}{*}{Yelp}
				&\multirow{9}{*}{Baselines}                                               
				& eAls    &  0.287                              & 0.289                              & 0.290                              & 0.167                              & 0.167                              & 0.169                              \\
				&& RCD     &
				0.454                              & 0.452                              & 0.447                              & 0.260                              & 0.257                              & 0.259                              \\
				&& iBPR    & 
				0.307                              & 0.295                              & 0.188                              & 0.180                              & 0.172                              & 0.108                              \\
				
				&& SPMF    &
				0.197                              & 0.192                              & 0.184                              & 0.104                              & 0.100                              & 0.097                              \\
				&& iGMF    & 
				0.499                              & 0.470                              & 0.396                              & 0.294                              & 0.276                              & 0.228                              \\
				&& iMLP    &
				
				\underline{0.573} & \underline{0.574} & \underline{0.438} & \underline{0.338} & \underline{0.338} & 0.246                              \\
				&& iNeuMF  & 
				
				0.566                              & 0.570                              & 0.435                              & 0.331                              & 0.334                              & \underline{0.247} \\ 
				&& OCFIF   & 
				0.260                              & 0.249                              & 0.203                              & 0.135                              & 0.129                              & 0.107                              \\ \cline{2-9}
				&Ours  
				&VRS-DWMoE\_8 
				& \textbf{0.608}    & \textbf{0.603}    & \textbf{0.602}    & \textbf{0.354}    & \textbf{0.358}    & \textbf{0.353}     \\ \cline{2-9}
				\rule{0pt}{8pt}&\multicolumn{2}{c|}{\makecell{Improvement percentages\textsuperscript{5}}}
				& 6.10\%                             & 5.10\%                             & 37.4\%                             & 4.70\%                             & 5.90\%                             & 42.9\%                                \\ \hline 
			\end{tabular}
		}
		\label{table:performance_comparison}
		\vspace{-2 mm}
	\end{table}
	
	\footnotetext{Improvement percentages over the best-performing baseline(s)}
	\vspace{-1 mm}
	\textbf{Experiment 1: Comparison with Baselines (for RQ1 and RQ2)}\\
	\noindent\textbf{Setting.} To answer \textbf{RQ1} and \textbf{RQ2}, we compare our proposed VRS-DWMoE (the number $n_e$ of experts is set to eight) with all eight baselines with a fixed data processing speed $s_p = 256$ and different data receiving speeds, where $s_r = 128$ and $s_r = 512$ indicate the underload scenario and overload scenario, respectively. \\
	\textbf{Result 1 (for RQ1).} \Cref{table:performance_comparison} shows the results of our proposed approach and eight baselines on all three datasets. In all the cases, VRS-DWMoE\_8 delivers the highest recommendation accuracies (marked with \textbf{bold} font), and the improvement percentages of VRS-DWMoE\_8 over the best-performing baselines (marked with \underline{underline}) are introduced in the last row for each dataset, ranging from 2.0\% to 37.4\% with an average of 8.5\% in terms of HR@10, and ranging from 1.9\% to 42.9\% with an average of 10.4\% in terms of NDCG@10.
	
	The superiority of VRS-DWMoE can be explained in two aspects: 1) VRS addresses user preference drift while capturing long-term user preferences by wisely complementing new data with sampled historical data, and 2) DWMoE better learns the heterogeneous user preferences and item characteristics with two MoEs, where each expert specializes in one underlying type of users or items.
	
	\noindent\textbf{Result 2 (for RQ2).} The superiority of our DWMoE is verified by the cases where $s_r = s_p$, i.e., $s_r = 256$, in \Cref{table:performance_comparison}. Specifically, in these cases, both our apporach and baselines utilize all the new data to train recommendation models, thus their recommendation accuracy only depends on their recommendation models. Therefore, the superiority of DWMoE is confirmed by the highest recommendation accuracy delivered by VRS-DWMoE in these cases. The reason for this superiority is that DWMoE not only learns heterogeneous user preferences and item characteristics with two dedicated MoEs, respectively, but also allows each of their experts to specialize in one underlying type of users or items.
	
\begin{figure}[t]
		\centering
		\begin{subfigure}[b]{.85\textwidth}
			\hfill
			\includegraphics[width=\textwidth]{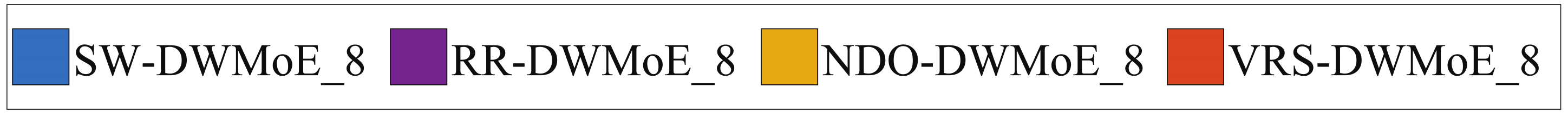}
			
		\end{subfigure}%
		
		\begin{subfigure}[b]{.42\textwidth}
			\centering
			\includegraphics[width=\textwidth]{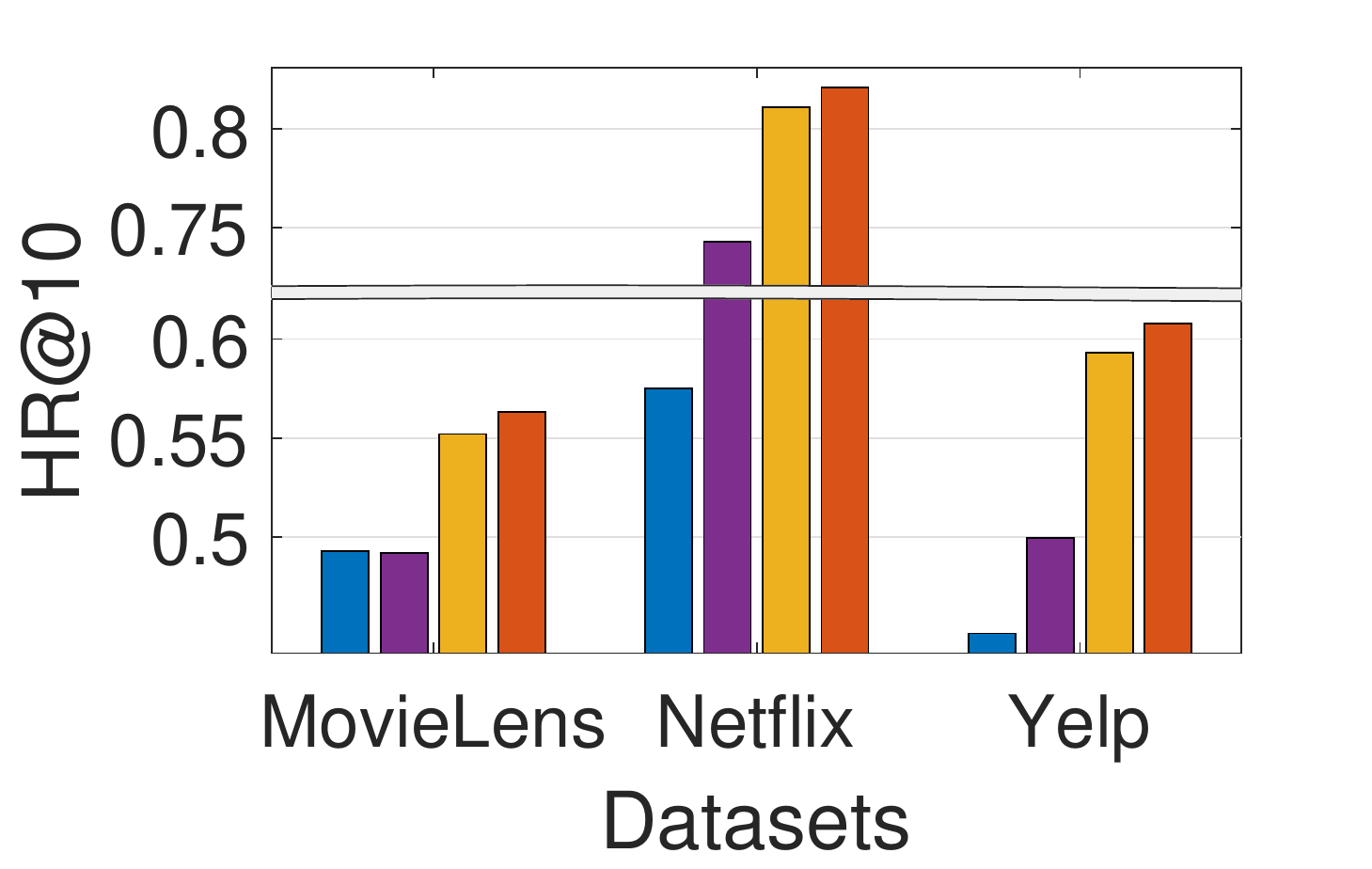}
			
		\end{subfigure}\hspace{6mm}
		\begin{subfigure}[b]{.42\textwidth}
			\centering
			\includegraphics[width=\textwidth]{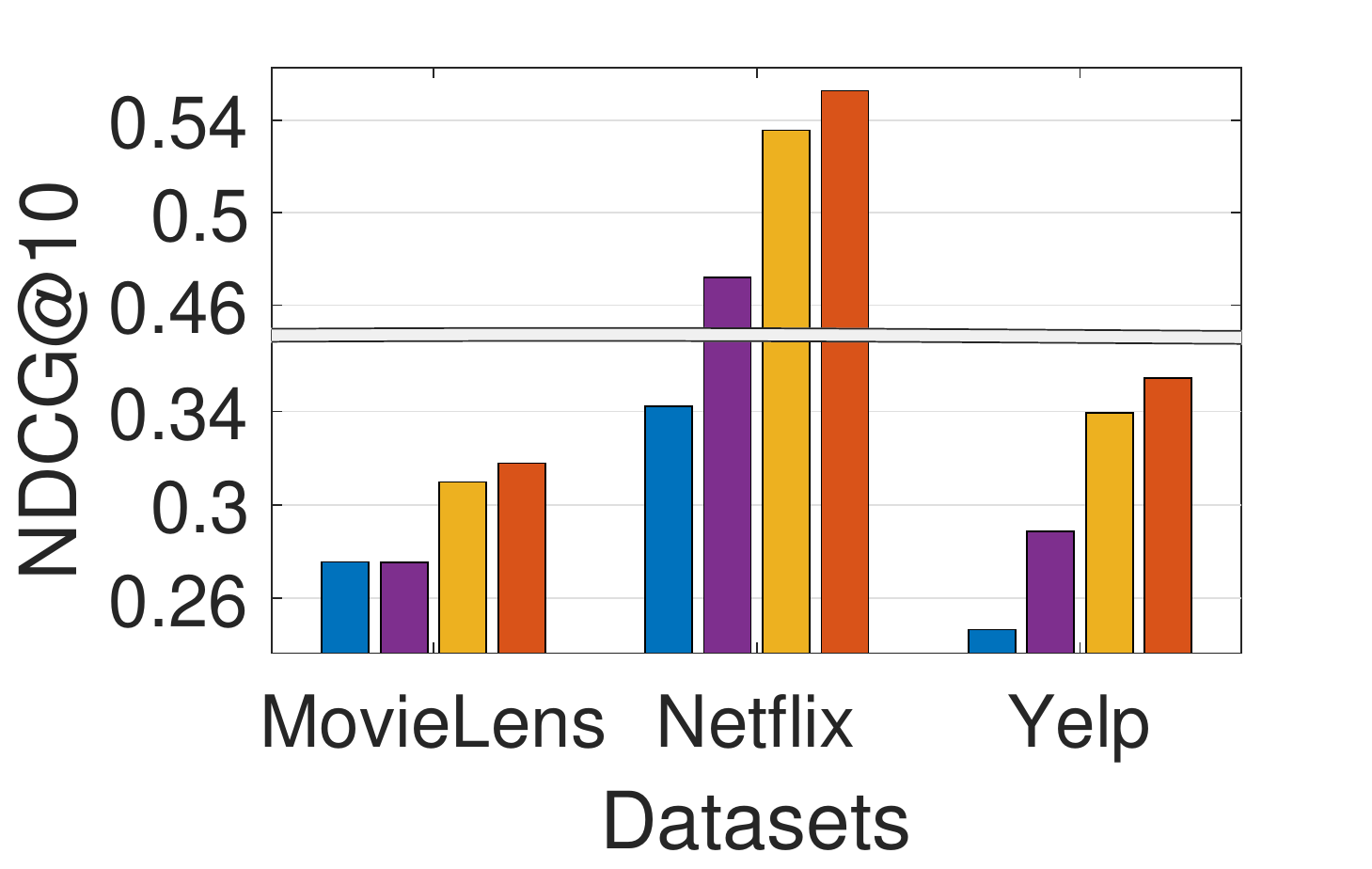}
			
		\end{subfigure}
		
		\caption{Performance of VRS. VRS outperforms all the other sampling methods.}
		\vspace{-5mm}
		
		\label{fig:sampling}
	\end{figure}

	\noindent\textbf{Experiment 2: Performance of VRS (for RQ3)}\\
	\textbf{Setting.} To answer \textbf{RQ3}, we replace our proposed  VRS with existing sampling methods, including New Data Only (NDO)~\cite{ocfif}, Reservoir-enhanced Random sampling (RR)~\cite{rmfk}, and Sliding Window (SW)~\cite{OWE}, for comparisons.  In this experiment, we report the results in the underload scenario only to save space while the results in the overload scenario are similar to the reported ones.\\
	\textbf{Result 3 (for RQ3).} As~\Cref{fig:sampling} illustrates, our proposed VRS outperforms all the other sampling methods. The improvements
	of VRS over the best-performing baseline, i.e., NDO, range from 1.2\%
	(on Netflix) to 2.0\% (on Yelp) with an average of 1.9\% in terms of HR@10, and range from
	3.2\% (on Netflix) to 4.3\% (on Yelp) with an average of 3.4\% in terms of NCDG@10. The effectiveness of VRS comes from wisely complementing new data with sampled historical data while guaranteeing the proportion of new data, and thus addressing user preference drift while capturing long-term user preferences.
	
	\noindent\textbf{Experiment 3: Impact of Number of Experts (for RQ4)}\\
	\noindent\textbf{Setting.} To answer \textbf{RQ4}, we compare the performance of VRS-DWMoE when equipped with different numbers ($n_e$) of experts, i.e., 2, 4, 6, and 8. In this experiment, we report the results in the overload scenario only to save space while the results in the underload are similar to the reported ones.
	\begin{figure}[t]
		\centering
		\begin{subfigure}[b]{.5\textwidth}
			\centering
			\includegraphics[width=\textwidth]{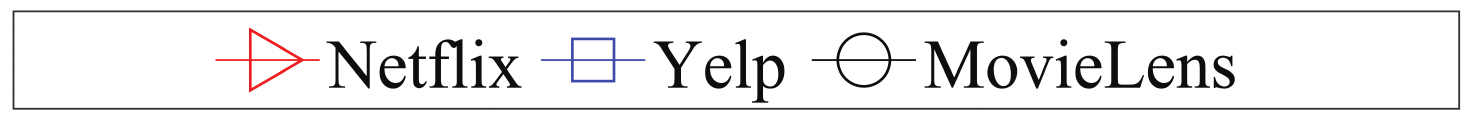}
			
		\end{subfigure}%
		
		\begin{subfigure}[b]{.42\textwidth}
			\centering
			\includegraphics[width=\textwidth]{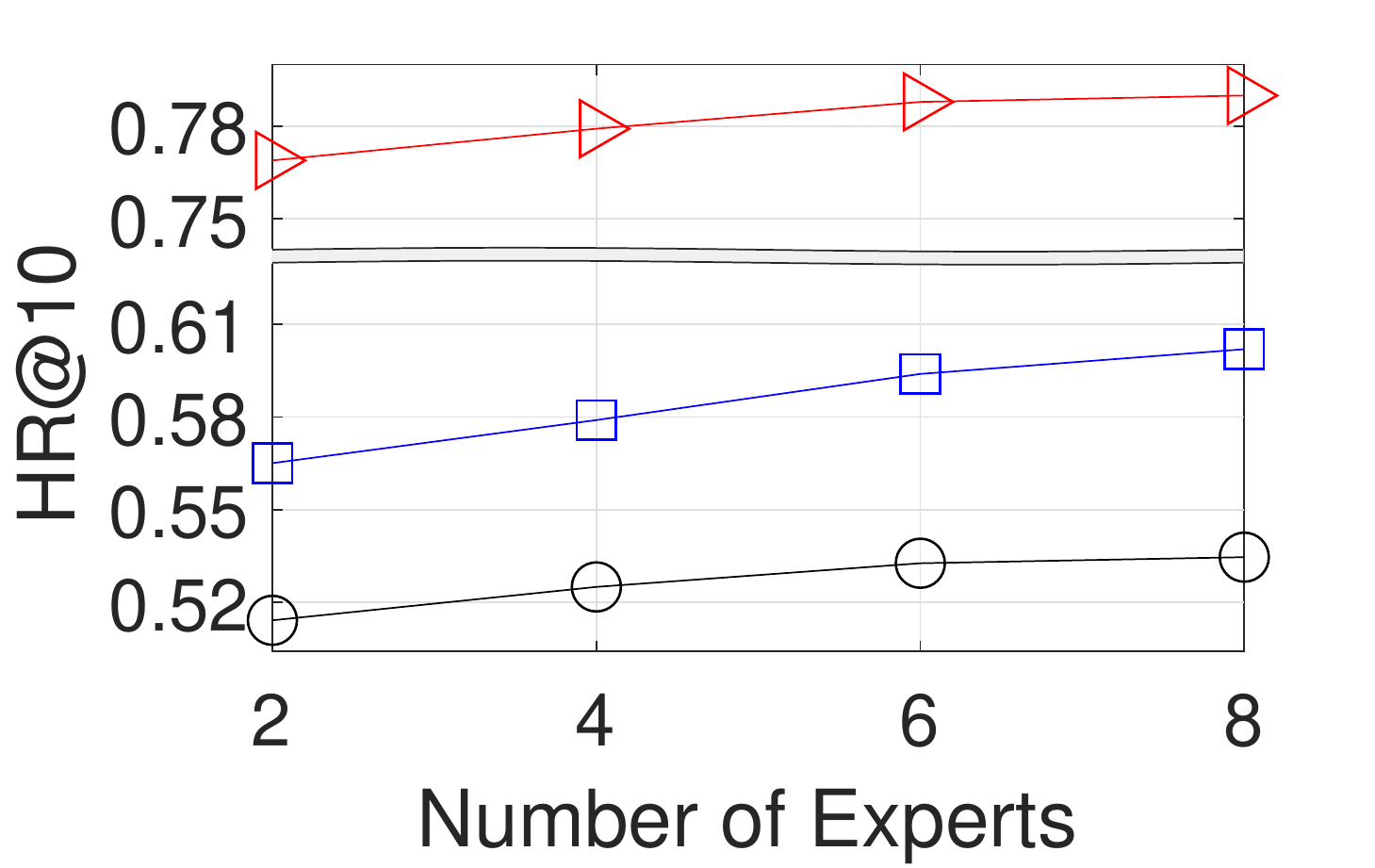}
		\end{subfigure} \hspace{6mm}
		\begin{subfigure}[b]{.42\textwidth}
			\centering
			\includegraphics[width=\textwidth]{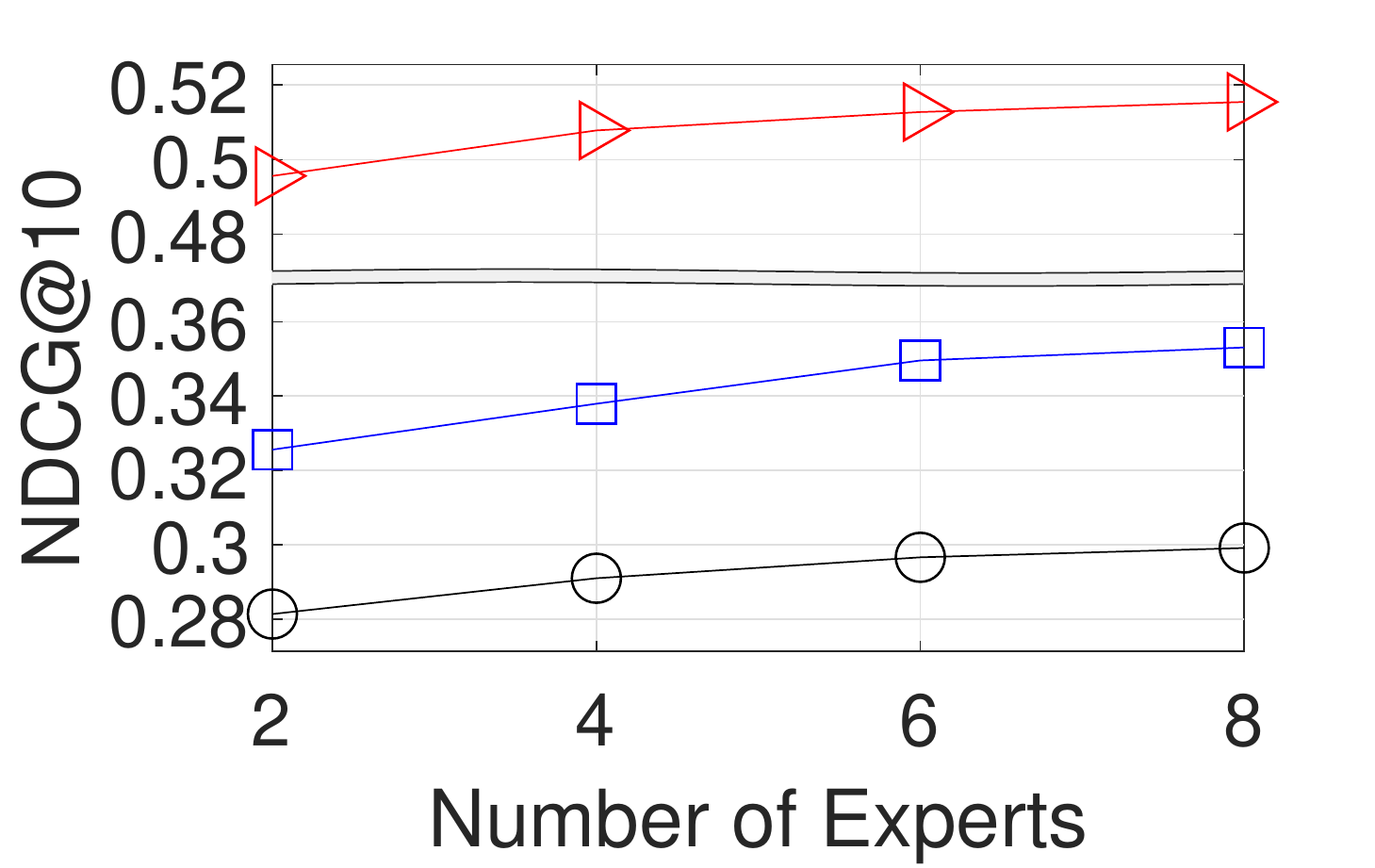}
		\end{subfigure}
		\caption{Impact of the number ($n_e$) of experts. VRS-DWMoE delivers higher recommendation accuracy with more experts.}
		
		\label{fig:number}
		
		\vspace{-4mm}
	\end{figure}
	
	\noindent\textbf{Result 4  (for RQ4).} As~\Cref{fig:number} illustrates, our proposed VRS-DWMoE delivers higher recommendation accuracy when equipped with more experts. The improvements of VRS-DWMoE equipped with eight experts over that equipped with two experts range from 2.7\%
	(on Netflix) to 6.5\% (on Yelp) with an average of 4.4\% in terms of HR@10, and range from
	4.0\% (on Netflix) to 8.4\% (on Yelp) with an average of 6.3\% in terms of NCDG@10. The reason for the superiority of more experts is that more experts better complement one another with their expertise to more effectively learn user preferences and item characteristics. \\
	
	\vspace{-6mm}
	
	\section{Conclusions}
	\vspace{-2mm}
	In this paper, we have proposed a Variational and Reservoir-enhanced Sampling based Double-Wing Mixture of Experts framework (VRS-DWMoE) for accurate streaming recommendations. We first propose VRS to wisely complement new data with sampled historical data to address user preference drift while capturing long-term user preferences. After that, with these sampled data, DWMoE learns heterogeneous user preferences and item characteristics with two MoEs, i.e., MoUE and MoIE, respectively, and then makes recommendations with learned preferences and characteristics. The superiority of VRS-DWMoE has been verified by extensive experiments. In the future, we will wisely utilize different numbers of experts in MoUE and MoIE and study more effective reservoir maintenance strategy for higher accuracy of streaming recommendations.
	\vspace{-6mm}
	\section{Acknowledgements}
	\vspace{-1mm}
	This work was partially supported by Australian Research Council Discovery Projects DP180102378 and DP210101810.
	
%	\bibliography
%	\clearpage
%\bibliographystyle{unsrt}  
%%% Remove comment to use the external .bib file (using bibtex).


\end{document}